\documentclass[reprint,aps,prb,amsmath,amssymb]{revtex4-1}

\usepackage[]{graphicx}
\usepackage[]{units}
\usepackage{times}
\usepackage{bm}
\usepackage[ulem=normalem]{changes}
\usepackage[normalem]{ulem}		
\usepackage{color}
\usepackage{xr} 				
\usepackage{cancel}
\usepackage{babel}
\definechangesauthor[color=blue]{SM}

\newcommand{\im}{\Im\textnormal{m}}

\newcommand{\kp}{\mathbf{k}}

\renewcommand{\emph}{\textit}

\begin{document}

\title{Excitonic insulator states in molecular functionalized atomically-thin semiconductors}
\author{Dominik Christiansen$^{1}$}
\author{Malte Selig$^{1}$}
\author{Mariana Rossi$^{2,3}$}
\author{Andreas Knorr$^{1}$}
\affiliation{$^{1}$Institut f\"ur Theoretische Physik, Nichtlineare Optik und Quantenelektronik, Technische Universit\"at Berlin, 10623 Berlin, Germany}
\affiliation{$^{2}$Max Planck Institute for the Structure and Dynamics of Matter, 22761 Hamburg, Germany}
\affiliation{$^{3}$Fritz Haber Institute of the Max Planck Society, 14195 Berlin, Germany}

\begin{abstract}
The excitonic insulator is an elusive electronic phase exhibiting a correlated excitonic ground state. Materials with such a phase are expected to have intriguing properties such as excitonic high-temperature superconductivity. However, compelling evidence on the experimental realization is still missing. Here, we theoretically propose hybrids of two-dimensional semiconductors functionalized by organic molecules as prototypes of excitonic insulators, with the exemplary candidate WS$_2$-F6TCNNQ. This material system exhibits an excitonic insulating phase at room temperature with a ground state formed by a condensate of interlayer excitons. To address an experimentally relevant situation, we calculate the corresponding phase diagram for the important parameters: temperature, gap energy, and dielectric environment. Further, to guide future experimental detection, we show how to optically characterize the different electronic phases via far-infrared to terahertz (THz) spectroscopy.
\end{abstract}

\maketitle


\textit{Introduction}: The excitonic insulator (EI) is a charge neutral, strongly interacting insulating phase that arises from spontaneous formation of excitons. The EI was first predicted theoretically in 1965 by L.V. Keldysh \cite{Keldysh1965} and systematically investigated by W. Kohn \cite{Jerome1967}. The insulating phase is anticipated to appear in semiconductors at thermodynamic equilibrium, as long as the predicted exciton binding energy exceeds the band gap. It presents an interesting platform for realizing many-body ground states of condensed bosons in solids. Although the concept has been known for almost 60 years, to date compelling experimental evidence of the excitonic insulator is still missing. Since the EI phase is predicted to host many novel properties, such as superfluidity \cite{Gupta2020,Perfetto2021}, excitonic high-temperature superconductivity \cite{Parmenter1970,Wang2019}, and exciton condensation, breakthroughs in finding this new class of insulators has attracted great attention over the last decades \cite{Varsano2017,Wang2019,Perfetto2019,Perfetto2020,Perfetto2020_2,Jia2020}.

There are a few materials, which are suspected to possess EI ground states in a solid state, namely 1T-TiSe$_2$, Ta$_2$NiSe$_5$ or TmSe$_{0.45}$Te$_{0.55}$ \cite{Cercellier2007,Monney2010,Wakisaka2009,Lu2017,Werdehausen2018,Bucher1991,Bronold2006,Fukutani2021}. However, it has been difficult to establish whether the EI state has been realized, because the expected electronic phase transition is accompanied by a structural phase transition, which makes it difficult to distinguish between EI and normal insulator or Peierls transition \cite{Rossnagel2002,Zenker2014,Jiang2018,Volkov2021}. Just recently, the discussion started about EIs in TMDC heterostructures  \cite{Du2017,Ataei2021,Ma2021,zhang2021correlated}.

\begin{figure}[t]
\begin{center}
\includegraphics[width=\linewidth]{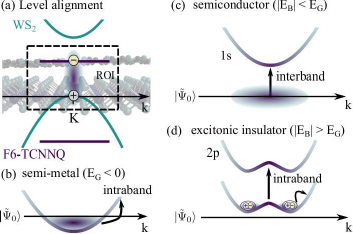}
\end{center}
\caption{(a) The HIOS consisting of a monolayer WS$_2$ and a layer of F6-TCNNQ molecules with level alignment in the free-particle picture. (b)-(d) The interlayer gap dispersion of the dashed boxed for the three possible electronic phases and the character of their optical excitation.}
\label{fig:sketch}
\end{figure}

In this Letter, we present a blueprint for the first realization of an EI based on the interlayer excitons of hybrid inorganic-organic systems (HIOS). HIOS is a growing field with increasing technological importance because it combines the best of two worlds: The strong light-matter interaction/tunability of the transition energies of organic molecules with the high carrier mobility of inorganic semiconductors \cite{Sun2019,Song2017,Koch2007,Fahlman2019}. For the construction of an EI, the low dielectric constant of the molecular lattices \cite{torabi2015strategy} and the strong localization of their electrons, due to an infinite effective mass, are excellent conditions for large exciton binding energies of HIOS interface excitons. In particular, the functionalization of atomically-thin semiconductors with organic molecules allows to choose a material combination with a band gap in an appropriate range \cite{Groom2016,Stuke2020} and the spatial indirect character of the exciton allows for static dipoles and thus unambiguous fine tuning via static Stark shifts \cite{Lorchat2021}. A the same time, the energy level tunability of the molecular layer has advantages over TMDC bilayer or heterobilayers, where the EI phase is still under discussion \cite{Du2017,Ataei2021,Ma2021,zhang2021correlated}.

Figure \ref{fig:sketch}(a) depicts the investigated HIOS with interlayer dispersion close to the K-point. The relevant electronic interlayer transition for the EI built up occurs between TMDC valence and molecular conduction band highlighted by the dashed line box. Due to the direct-gap character of the heterostructure dispersion, we also circumvent the formation of Peierls charge density waves. Depending on the interlayer band gap (adjustable also by tuning of an applied voltage) three different electronic phases can be expected: semi-metal, semiconductor, and EI, cp. Fig. \ref{fig:sketch}(b)-(d). We will show that all three phases can be distinguished by far-infrared/THz absorption: The semi-metal phase exhibits a vanishing or negative band gap with a free electron gas as ground state, cf. Fig. \ref{fig:sketch}(b). The optical response will then be determined by the Drude model known from metals \cite{drude1900elektronentheorie}. The semiconducting phase, cf. Fig. \ref{fig:sketch}(c), is determined by an exciton binding energy smaller than the free-particle band gap resulting in a filled valence band as ground state. The optical response is given as Lorentz response, however due to the interlayer character of the transition with considerably small oscillator strength \cite{Gillen2021,Deilmann2018}. In contrast, the excitonic insulating phase, cf. Fig. \ref{fig:sketch}(d), occurs if the expected exciton binding energy is larger than the band gap. We will exploit that the far-infrared response of such EIs are characterized by transitions in the exciton ladder \cite{Kira2006,bhattacharyya2014magnetic} of condensed ground state excitons to higher bound states.

\begin{figure}[t]
\begin{center}
\includegraphics[width=\linewidth]{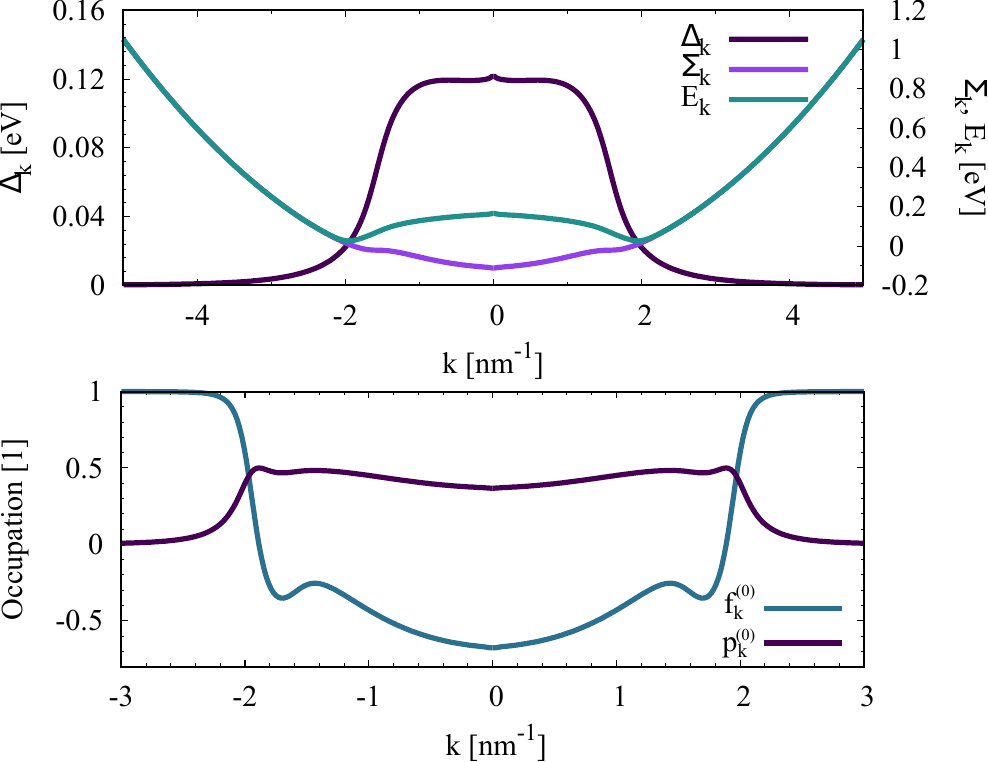}
\end{center}
\caption{(a) Numerical solution of the ordering parameter $\Delta_{\kp}$ and gap dispersion $\Sigma_{\kp}$ as function of wave number at room temperature. Both quantities constitute the Bogoliubov dispersion $E_{\kp}$. (b) The EI exhibits a non-vanishing coherence and an intrinsic inversion. The coherence and occupations peak in the pocket of the Bogoliubov dispersion.} 
\label{fig:groundState}
\end{figure}

In the following, even without applied electric field for energy band fine tuning, we predict an EI from a hybrid structure of a WS$_2$ monolayer and a self-assembled layer of F6-TCNNQ molecules, cf. Fig. \ref{fig:sketch}(a). The organic molecules form a periodic two-dimensional lattice, where we investigate the realistic ratio of one molecule per 16 WS$_2$ unit cells due to the weak molecule-WS$_2$ interaction\cite{Park2021}. From first-principles calculations we find a type-II band alignment heterostructure with an interlayer gap between WS$_2$ valence band and F6-TCNNQ lowest unoccupied molecular orbital of $E_G=\unit[0.12]{eV}$. The Kohn-Sham energy levels of F6-TCNNQ and WS$_2$ were obtained with the range-separated HSE06 hybrid functional \cite{krukau2006influence}, as implemented in the FHI-aims program \cite{blum2009ab,ren2012resolution,levchenko2015hybrid} and using standard \glqq intermediate\grqq ~ settings \cite{Park2021,wang2019modulation}. The naively calculated interlayer exciton binding energy, using well documented methods \cite{malic2018dark}, with completely filled valence band as ground state is $E_B=\unit[0.14]{eV}$. This is on the same order of magnitude as the band gap of \unit[0.12]{eV}, indicating the possibility of forming a strongly correlated insulating phase to minimize energy. Based on this observation, we develop in the following the description of the EI for the introduced hybrid and calculate its optical response. In particular, we show that the excitonic condensate can be stable up to room temperature under the proper choice of substrate.

\textit{Ground state of EI}: To calculate the EI ground state, we diagonalize the field-independent part of the mean-field Hamiltonian (cf. appendix) by a Bogoliubov transformation \cite{Bogoljubov1958,Valatin1958} with $\alpha^{\dagger}_{\mathbf{k}}=u^{*}_{\mathbf{k}}v^{\dagger}_{\mathbf{k}} - w^{*}_{\mathbf{k}}c^{\dagger}_{\mathbf{k}}$ and $\beta^{\dagger}_{\mathbf{k}}=w_{\mathbf{k}}v^{\dagger}_{\mathbf{k}} + u_{\mathbf{k}}c^{\dagger}_{\mathbf{k}}$ \cite{Jerome1967,Keldysh1968,Comte1982}. The fermionic operators $\alpha^{\dagger}_{\mathbf{k}}$ and $\beta^{\dagger}_{\mathbf{k}}$ create an electron in a linear combination of valence $v^{(\dagger)}_{\mathbf{k}}$ and conduction $c^{(\dagger)}_{\mathbf{k}}$ bands in analogy to the Bogoliubov particle operators from BCS superconductivity theory. The diagonalized Hamiltonian reads $H=\sum_{n=\{\alpha,\beta\},\mathbf{k}}E_{n,\mathbf{k}}n^{\dagger}_{\mathbf{k}}n^{\mathstrut}_{\mathbf{k}}$ with the hybridized bands $E_{\alpha/\beta,\mathbf{k}}=(\tilde{\varepsilon}_{c,\mathbf{k}}+\tilde{\varepsilon}_{v,\mathbf{k}})/2\mp\sqrt{\Sigma^2_{\mathbf{k}}+\Delta^2_{\mathbf{k}}}$. The excitation spectrum of the new quasi-particles corresponds to the Bogoliubov dispersion $E_{\mathbf{k}}=\sqrt{\Sigma^2_{\mathbf{k}}+\Delta^2_{\mathbf{k}}}$, where the gap dispersion is defined as $\Sigma_{\mathbf{k}}=(\tilde{\varepsilon}_{c,\mathbf{k}}-\tilde{\varepsilon}_{v,\mathbf{k}})/2$. The quantity $\Delta_{\mathbf{k}}$ is determined via the transcendental gap equation \cite{Kozlov1965,Sabio2010,Stroucken2011}
\begin{align}
\Delta_{\mathbf{k}}&=\frac{1}{2}\sum_{\mathbf{k'}} V_{\mathbf{k-k'}} \frac{\Delta_{\mathbf{k}'}}{\sqrt{\Sigma^2_{\mathbf{k}'}+\Delta^2_{\mathbf{k}'}}} \left(f_{\alpha,\mathbf{k}'}-f_{\beta,\mathbf{k}'} \right)  \label{eq:gapEq} \\
2\Sigma_{\mathbf{k}}&=\varepsilon_{c,\mathbf{k}}-\varepsilon_{v,\mathbf{k}}+\sum_{\mathbf{k}'} \mathcal{V}^{\text{mol}}_{\mathbf{k-k'}} \left(u^2_{\mathbf{k}'}f_{\beta,\mathbf{k}'}+w^2_{\mathbf{k}'}f_{\alpha,\mathbf{k}'} \right) \nonumber \\
&-\sum_{\mathbf{k}'} \mathcal{V}^{\text{WS}_2}_{\mathbf{k-k'}} \left(u^2_{\mathbf{k}'}f_{\alpha,\mathbf{k}'}+w^2_{\mathbf{k}'}f_{\beta,\mathbf{k}'} \right) \;. \label{eq:GapDispersion}
\end{align}
The coupled system of equations determines $\Delta_{\mathbf{k}}$, $u_{\mathbf{k}}$, and $w_{\mathbf{k}}$ via temperature and band gap. The quantity $\Delta_{\mathbf{k}}$ can be identified as ordering parameter determining the phase of the heterostructure. A finite value accounts for a finite probability to create electron-hole pairs, which designates the excitonic instability. In the limit $\Delta_{\kp}\rightarrow 0$ the dipolar excitonic insulator \cite{zimmermann2007exciton,schindler2008analysis,gu2021dipolar} converts to the conventional phases of semiconductor or semi-metal depending on the band gap. In Eq. \eqref{eq:gapEq} and \eqref{eq:GapDispersion} we defined the matrix element $\mathcal{V}^{l}_{\mathbf{k-k'}}=V_{\mathbf{0}}^{l} - V_{\mathbf{k-k'}}^{l}-V_{\mathbf{0}}+V_{\mathbf{k-k'}}$ with intralayer Coulomb potential $V^l_{\kp}$ ($l=\{\text{molecule,WS}_2\}$), and interlayer potential $V_{\kp}$\cite{Ovesen2019} (cf. appendix). Equation \eqref{eq:gapEq} and \eqref{eq:GapDispersion} depend on the occupations of the hybridized bands $f_{n,\mathbf{k}}=\langle n^{\dagger}_{\mathbf{k}}n^{\mathstrut}_{\mathbf{k}}\rangle$ with $n=\{\alpha,\beta\}$. The temperature $T=\unit[0]{K}$ limit of the gap equation, i.e. $f_{\alpha,\mathbf{k}}-f_{\beta,\mathbf{k}}=1$, can also be obtained from a minimization of the energy \cite{Sabio2010,Stroucken2011}. Clearly, the magnitude of the ordering parameter $\Delta_{\mathbf{k}}$ depends on band gap and temperature, which enters via the Fermi functions $f_{\alpha/\beta,\mathbf{k}}$ of the hybridized bands $E_{\alpha/\beta,\mathbf{k}}$. The chemical potential is chosen such that the charge density is a conserved quantity as function of temperature and band gap. In our case, we consider a charge neutral structure, i.e. the density of holes in WS$_2$ equals the density of electron in the molecule. Because the ordering parameter enters also in the gap dispersion $\Sigma_{\mathbf{k}}$ via the Fermi functions $f_{n,\mathbf{k}}(E_{n,\kp})$, both quantities have to be solved simultaneously. If not stated otherwise  we use an hBN substrate ($\epsilon=4.5$) entering the Coulomb potential and a surrounding of air ($\epsilon=1$) for the numerical evaluation.

\begin{figure}[t]
\begin{center}
\includegraphics[width=\linewidth]{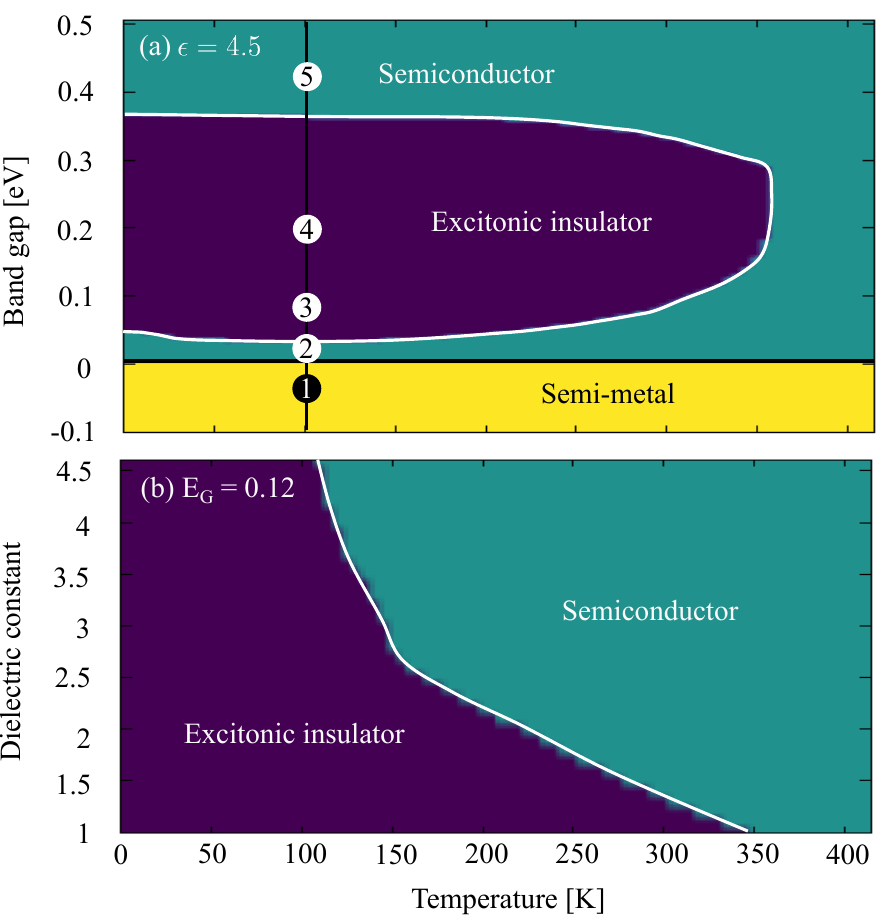}
\end{center}
\caption{(a) Phase diagram of a WS$_2$-F6TCNNQ stack on hBN and air surrounding. The EI phase is seperated from the semi-metal by the semiconducting phase. The numbers denote the position for the absorption spectra in Fig. \ref{fig:absorption}. Number 3 corresponds to zero applied field. (b) Phase diagram of the WS$_2$-F6TCNNQ stack on hBN and altering supstrate, characterized by the dielectric constant. Environments with low dielectric constant are favourable to stabilize the excitonic insulating phase at higher temperature.} 
\label{fig:phasediagram}
\end{figure}

Figure \ref{fig:groundState}(a) displays the numerical solution of $\Delta_{\mathbf{k}}$ and $\Sigma_{\mathbf{k}}$ at room temperature, and the resulting Bogoliubov dispersion $E_{\mathbf{k}}$ with the original band gap of $E_G=\unit[0.12]{eV}$. The ordering parameter $\Delta_{\mathbf{k}}$ is symmetric to $k=0$ and displays a monotonous decrease with the wave number, comparable to the gap function of s-wave superconductors \cite{Bardeen1957,Tinkham2004,Reyes2016}. Together with $\Sigma_{\mathbf{k}}$ it yields a sombrero-like Bogoliubov dispersion. We stress that Fig. \ref{fig:groundState}(a) displays the results for \unit[300]{K} already suggesting that the excitonic condensate is stable up to room temperature.
Finally, Fig. \ref{fig:groundState}(a) displays also the gap dispersion $\Sigma_{\mathbf{k}}$. For the discussion of the gap dispersion, it is convenient to introduce the spontaneously forming inversion  $f^{(0)}_{\mathbf{k}}\equiv f^{(0)}_{v,\mathbf{k}}-f^{(0)}_{c,\mathbf{k}}=\Sigma_{\mathbf{k}}/\sqrt{\Sigma^2_{\mathbf{k}}+\Delta^2_{\mathbf{k}}}(f_{\alpha,\mathbf{k}}-f_{\beta,\mathbf{k}})$ and the macroscopic coherence $p^{(0)}_{\mathbf{k}}=\langle v^{\dagger}_{\mathbf{k}}c^{\mathstrut}_{ \mathbf{k}}\rangle^{(0)}=\frac{1}{2}\Delta_{\mathbf{k}}/\sqrt{\Sigma^2_{\mathbf{k}}+\Delta^2_{\mathbf{k}}}(f_{\alpha,\mathbf{k}}-f_{\beta,\mathbf{k}})$ forming without external source and plotted in Fig. \ref{fig:groundState}(b). Here, in contrast to a conventional semiconductor ($\Delta_{\kp}= 0$) for $\Delta_{\kp}\neq 0$ the ground state coherence $p^{(0)}_{\kp}$ has a non-vanishing value and the occupation inversion deviates from unity clarifying why $\Delta_{\mathbf{k}}$ is referred to as ordering parameter. Due to the Hartree-Fock renormalizations the gap dispersion $\Sigma_{\kp}$ turns negative, cf. Fig. \ref{fig:groundState}(a). This leads to an intrinsic inversion $f^{(0)}_{\mathbf{k}}<0$ close to the band extremum. From Fig. \ref{fig:groundState}(b) we see that the ground state coherence and occupations peak within the pockets of the Bogoliubov dispersion, which we can identify with the Fermi wave number $k_F$.

\begin{figure}[t]
\begin{center}
\includegraphics[width=\linewidth]{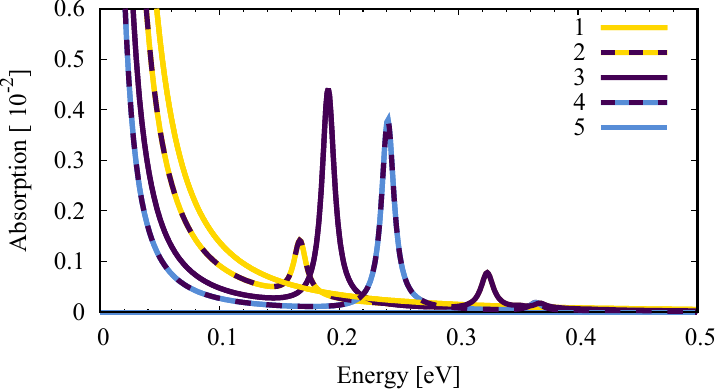}
\end{center}
\caption{Absorption spectrum at \unit[100]{K} along the dotted line in Fig. \ref{fig:phasediagram}(a). The Drude response in the semi-metal phase is modulated by a $p$-exciton Rydberg series for the EI phase in the terahertz regime. In the semiconducting phase the terahertz response vanishes.}
\label{fig:absorption}
\end{figure}

\textit{Phase diagram}: Depending on the two external parameters temperature and band gap, tunable by static electric fields \cite{Deilmann2018,Lorchat2021}, affecting the ordering parameter $\Delta_{\mathbf{k}}\neq 0$ we can expect three different electronic phases: EI, semiconductor, and semi-metal, cf. Fig. \ref{fig:sketch}(b)-(d). While the EI phase is present for a finite ordering parameter $\Delta_{\mathbf{k}}$, the other two phases reveal a vanishing $\Delta_{\mathbf{k}}$. However, semi-metal and semiconductor can be distinguished via the inversion $f^{(0)}_{\mathbf{k}}$: While for a conventional semiconductor the inversion is one in the ground state, the value is smaller than one for a semi-metal reflecting the presence of a free electron gas. Figure \ref{fig:phasediagram}(a) shows the calculated phase diagram of a WS$_2$-F6TCNNQ stack on hBN substrate as function of temperature and band gap. As guidance we include the coexistence lines between the different phases. We find that the excitonic insulating phase is stable for temperatures up to \unit[350]{K} underlining the ability of our proposed structure as high-temperature EI. The EI phase appears in the band gap range of \unit[0.04]{eV} to \unit[0.38]{eV}. We see that by increasing the band gap the heterostructure enters its semiconducting phase. When decreasing the band gap we approach the semi-metal limit. Interestingly, we find no coexistence line between EI and semi-metal, but the heterostructure traverses the semiconducting phase again. This results from a fast decrease of the ordering parameter, vanishing prior to a negative band gap. The corresponding separating area between EI and semi-metal could be understood as an excited semiconducting phase ($\Delta_{\mathbf{k}}=0$, $E_{\mathbf{k}}=\tilde{\varepsilon}_{c,\mathbf{k}}-\tilde{\varepsilon}_{v,\mathbf{k}}>0$ as in the semiconducting phase but with $f_{c,\mathbf{k}}\neq 0$, $f^{(0)}_{\kp}<1$). To tune the band gap over the full energy range, which is discussed in Fig. \ref{fig:phasediagram}(a), a maximum applied voltage of \unit[0.5]{V} between molecule and inorganic semiconductor is necessary. Finally, we see a sublimation line at high enough temperature -- a direct transition from semiconductor to semi-metal. The billiard balls in Fig. \ref{fig:phasediagram}(a) are discussed later together with Fig. \ref{fig:absorption}.

We stress that the phase diagram strongly depends on the dielectric environment influencing the interlayer Coulomb potential \cite{Ovesen2019}: In Fig. \ref{fig:phasediagram}(b) we show the phases for the fixed original band gap of $E_G=\unit[0.12]{eV}$ with altering temperature and dielectric surrounding. The WS$_2$-F6TCNNQ stack is still placed on top of an hBN substrate, as in all calculations before, but the dielectric constant of a top layer is changed from vacuum to hBN encapsulation: The stability of the condensate for higher temperature rapidly decreases with increasing dielectric constant. Obviously, a dielectric environment with low mean dielectric constant is favourable to stabilize the EI phase at high temperatures.

\textit{Optical response}: Based on the ground state calculations, we can now calculate the frequency-dependent absorption coefficient via a self-consistent coupling of material and wave equation \cite{Knorr1996,Malic2013,Katsch2020}. The linear absorption with respect to the dynamical field $\mathbf{E}(t)$ is determined by the susceptibility described by the microscopic polarization $p_{\mathbf{k}}$ and occupations $f_{\mathbf{k}}$. Both quantities are expanded up to first order in the exciting electric field \cite{Stroucken2011,Stroucken2015}. The initial conditions arise from the ground state (initially calculated $p^{(0)}_{\mathbf{k}}$ and $f^{(0)}_{\mathbf{k}}$). The dynamical correction to first order in $\mathbf{E}(t)$ is denoted by $p^{(1)}_{\mathbf{k}}$ and $f^{(1)}_{\mathbf{k}}$. The Heisenberg equation of motion for the microscopic polarization $p^{(1)}_{\mathbf{k}}$ can be diagonalized exploiting the Bogoliubov-Wannier equation \cite{Stroucken2015}. After diagonalization, based on a 1$s$ ground state, it describes far-infrared transitions to higher lying excitonic states $\mu$ with their excitonic energy $E_{\mu}$ and wave function $\varphi_{\mu,\mathbf{k}}$. Consequently, the microscopic polarization can be expressed as excitonic polarization $p_{\mu}=\sum_{\mathbf{k}}\varphi^*_{\mu,\mathbf{k}}p^{(1)}_{\mathbf{k}}$ \cite{Kira2006,Katsch2020_B,dong2021direct} (details in appendix). A direct solution of the Heisenberg equations of motion for excitonic polarization and excited occupation in frequency domain yields the susceptibility, detectable in far-infrared/THz experiments:
\begin{align}
\chi(\omega)=-\frac{1}{\epsilon_0}\sum_{\mu}\frac{\mathbf{d}_{\mu}\otimes\mathbf{d}_{\mu} + \mathbf{j}_{\mu}\otimes\mathbf{j}_{\mu}}{\hbar\omega-E_{\mu}+i\gamma} +\frac{e_0^2}{\epsilon_0\hbar}\sum_{\kp} \frac{\mathbf{v}_{\mathbf{k}}\otimes\nabla_{\kp}f^{(0)}_{v,\kp}}{\omega^2+i\gamma\omega/\hbar}  \label{eq:susci}
\end{align}
with the particle velocity $\mathbf{v}_{\kp}=\hbar\kp/m$, elementary charge $e_0$, and valence electron mass $m$. The excitonic \textit{interband} matrix element reads $\mathbf{d}_{\mu}=\mathbf{d}\sum_{\mathbf{k}} f^{(0)}_{\mathbf{k}}\varphi^*_{\mu,\mathbf{k}}$ with electronic dipole moment $\mathbf{d}$. The second summand is driven by the excitonic \textit{intraband} matrix element $\mathbf{j}_{\mu}=e_0\cdot\sum_{\mathbf{k}}\varphi^*_{\mu,\mathbf{k}}\nabla_{\mathbf{k}}p^{(0)}_{\mathbf{k}}$. Both contributions exhibit resonances at the exciton energy $E_{\mu}$. For an EI with $s$-symmetric ground state the first excited exciton is of $p$-symmetry, that we can expect the 1$s$-2$p$ transition as the energetically lowest resonance \cite{Kira2006,bhattacharyya2014magnetic}. For a $p$-excited state ($\mu=p$ in Eq. \eqref{eq:susci}), the interband source vanishes because of its uneven parity. In contrast, the intraband source $\mathbf{j}_{\mu}$ is finite. When entering the semiconducting phase, the Bogoliubov-Wannier equation yields as lowest excited state 1$s$ excitons. In this phase, the optical source is of interband nature from a fully occupied valence band as ground state, cf. Fig. \ref{fig:sketch}(c). In contrast the intraband source vanishes due to symmetry reasons. Additionally, now holds $f^{(0)}_{v,\kp}=1$ (full valence band) that the third term in Eq. \eqref{eq:susci}, which resembles to the conductivity tensor of a plasma, vanishes. Finally, for zero or negative band gap the heterostructure is in its semi-metallic phase and the optical matrix elements $\mathbf{d}_{\mu}$ and $\mathbf{j}_{\mu}$ are zero due to the corresponding ground state coherence and inversion. The optical response is now solely described by the third term in Eq. \eqref{eq:susci}. We can perform a partial integration to bring the third term in the susceptibility into the form $\chi_{\text{Drude}}(\omega)=\omega_{pl}^2/(\omega^2+i\gamma\omega/\hbar)$, which corresponds to the plasma response in a Drude model for a free electron gas, cf. Fig. \ref{fig:sketch}(b). It stems from intraband transitions of the microscopic occupation with the plasma frequency $\omega^2_{pl}=e^2_0\sum_{\mathbf{k}}f^{(0)}_{v,\mathbf{k}}/\epsilon_0m$. Also for the EI phase the Drude response is present, since the ground state valence band occupation $f^{(0)}_{v,\mathbf{k}}$ exhibits a Fermi edge. In order to account for the broadening of the response, we include a phenomenological dephasing $\gamma$ \cite{Selig2016,Christiansen2017,helmrich2021phonon}. A detailed investigation of the exciton lifetime in the insulating phase would require an evaluation of exciton-phonon interaction into the HIOS Bloch equations.

Figure \ref{fig:absorption} displays the calculated absorption along the enumerated line in Fig. \ref{fig:phasediagram}(a). In the semi-metal phase (dot 1) with overlapping valence and conduction band we observe the well-known Drude response for a free electron gas. Due to the infinite mass of the conduction band electrons in the molecules the response stems solely from the half filled WS$_2$ valence band. Opening the gap (dot 2) an additional feature arises, which stems from the intraband matrix element in Eq. \eqref{eq:susci}. The rising macroscopic coherence $p^{(0)}_{\mathbf{k}}$ is convoluted with the excitonic wave function, which interpolates between $s$ and $p$ state. When entering the EI phase (dot 3) the Drude response is modulated by a $p$-excitonic Rydberg series in the far-infrared/THz regime stemming from the transitions from 1$s$ ground state to $p$-excited states. Since the optical source of the observed 1$s$-2$p$ transition is of intraband nature, it has sufficient oscillator strength (dot 3 and 4) to be observed in optical experiments. The oscillator strength decreases as a function of the exciton number due to a decrease of the integral in Eq. \eqref{eq:susci}. For a further gap opening we observe a blue shift of the exciton and a decreasing oscillator strength since the ordering parameter and the connected ground state functions $f^{(0)}_{\mathbf{k}}$ and $p_{\mathbf{k}}^{(0)}$ decrease. Finally, for a band gap of \unit[0.4]{eV} (dot 5) the wave function of the lowest excited state changes to $s$-symmetry, that the intraband source vanishes, marking the transition to the semiconducting phase. The interlayer exciton is now driven by the interband dipole matrix element. Because of the large detuning of intra- and interlayer exciton, the hybridization is small yielding an extremely small electronic dipole element \cite{Lorchat2021,Deilmann2018,Gillen2021}. Also the Drude response vanishes due to $f^{(0)}_{\mathbf{k}}=1$. Therefore, the interlayer exciton is not observable in absorption, the far-infrared/THz response vanishes, and the heterostructure becomes transparent for these wavelength. For the situation at room temperature a similar picture emerges characterized by a generally weaker oscillator strength since the ground states acting as sources are less populated.

\textit{Conclusion:} In summary, we propose HIOS as a candidate for the realization of an EI exploiting interlayer excitons. For this, the molecule decoration of an atomically-thin semiconductor is chosen in a way that the interlayer band gap lies in the range of the binding energy of the interlayer excitons. Additional static fields can be used for fine tuning or to induce different electronic phases. The occurring optical far-infrared/THz response can be used to characterize the electronic phases: While the conventional interlayer semiconductor exhibits a $s$-like Rydberg series with minimal oscillator strength, the EI's Rydberg series has $p$-character with strong oscillator strength including a Drude-like response from the ground state occupation. We expect that our results trigger new studies on high-temperature exciton condensation and an experimental realization of the predicted room temperature EI could lead to new optoelectronic applications.


\vspace{5mm}
We thank Florian Katsch, Manuel Katzer, Lara Greten (TU Berlin) and Kiril Bolotin (FU Berlin) for fruitful discussions. We acknowledge financial support from the Deutsche Forschungsgemeinschaft (DFG) through SFB 951 (D.C., M.S., A.K.) Projektnummer 182087777. D.C. thanks the graduate school Advanced Materials (SFB 951) for support. 

\appendix
\onecolumngrid

\section{Hamiltonian}
Due to the periodic arrangement of the molecules, the molecular operators of the lowest unoccupied orbital $c$ of molecule $\nu$ can be transformed to a Bloch basis $c_{\mathbf{k}}=\sum_{\nu}\exp(-i\mathbf{k}\cdot\mathbf{R}_{\nu})c_{\nu}/\sqrt{N_{\text{m}}}$ with $N_{\text{m}}$ the number of molecular unit cells and wave vector $\mathbf{k}$ \cite{Slater1934,Kaplan1976,Specht2018}. At the same time, valence band electrons in the TMDC layer are described by operators $v_\mathbf{k}$. In this notation, we can construct a many-particle Hamiltonian, which is parameterized from electronic structure \textit{ab initio} calculations:
\begin{align}
H&=\sum_{\lambda,\mathbf{k}} \varepsilon_{\lambda,\mathbf{k}} \lambda^{\dagger}_{\mathbf{k}}\lambda^{\mathstrut}_{\mathbf{k}} +ie_0\sum_{\lambda,\mathbf{k}}\mathbf{E}(t)\cdot(\nabla_{\mathbf{k}}\lambda^{\dagger}_{\mathbf{k}})\lambda^{\mathstrut}_{\mathbf{k}} + \hbar\sum_{\mathbf{k}}\Omega_{\mathbf{k}}(t)\left( v^{\dagger}_{\mathbf{k}}c^{\mathstrut}_{\mathbf{k}} + c^{\dagger}_{\mathbf{k}}v^{\mathstrut}_{\mathbf{k}}  \right) + \frac{1}{2}\sum_{\substack{\lambda,\lambda',\nu,\nu' \\ \mathbf{k,k',q}}}V^{\lambda\nu\nu'\lambda'}_{\mathbf{k,k',q}}~ \lambda^{\dagger}_{\mathbf{k+q}}\nu^{\dagger}_{\mathbf{k'-q}}\nu'^{\mathstrut}_{\mathbf{k'}}\lambda'^{\mathstrut}_{\mathbf{k}} .
\end{align}
The first term describes the free kinetic energy. The single-particle energies are $\varepsilon_{c,\mathbf{k}}=E_G$ and $\varepsilon_{v,\mathbf{k}}=\hbar^2\mathbf{k}^2/2m$ with effective hole mass $m$. The second and third term describe intra- and interband transitions, respectively. The last term includes Coulomb interaction. To account for inter- and intraband renormalization effects and exciton formation we can restrict the band indices to the combinations of: all band indices correspond to valence or conduction band and to even numbers of valence and conduction band indices. Uneven numbers of band indices would describe Meitner-Auger processes. However, since valence and conduction band are in different layers, the Meitner-Auger process would require a large wave function overlap to significantly contribute and can therefore be neglected \cite{dong2021observation}. Together with a random phase approximation for electronic occupations $\langle\lambda^{\dagger}_{\mathbf{k}}\lambda^{\mathstrut}_{\mathbf{k}'}\rangle\rightarrow\langle\lambda^{\dagger}_{\mathbf{k}}\lambda^{\mathstrut}_{\mathbf{k}'}\rangle\delta_{\mathbf{k,k'}}$ we obtain the Hartree-Fock Hamiltonian
\begin{align}
H&=\sum_{\mathbf{k}}\left(\varepsilon_{c,\mathbf{k}} + \sum_{\mathbf{k'}} V^{\text{mol}}_{\mathbf{0}} f_{c,\mathbf{k'}}  + \sum_{\mathbf{k'}} V_{\mathbf{0}} f_{v,\mathbf{k'}} - \sum_{\mathbf{k'}} V^{\text{mol}}_{\mathbf{k-k'}} f_{c,\mathbf{k'}} - \sum_{\mathbf{k'}} V_{\mathbf{k-k'}} f_{v,\mathbf{k'}}\right) c^{\dagger}_{\mathbf{k}}c^{\mathstrut}_{\mathbf{k}} \nonumber \\
&+\sum_{\mathbf{k}}\left(\varepsilon_{v,\mathbf{k}} + \sum_{\mathbf{k'}} V^{\text{WS}_2}_{\mathbf{0}} f_{v,\mathbf{k'}}  + \sum_{\mathbf{k'}} V_{\mathbf{0}} f_{c,\mathbf{k'}} - \sum_{\mathbf{k'}} V^{\text{WS}_2}_{\mathbf{k-k'}} f_{v,\mathbf{k'}} - \sum_{\mathbf{k'}} V_{\mathbf{k-k'}} f_{c,\mathbf{k'}}\right) v^{\dagger}_{\mathbf{k}}v^{\mathstrut}_{\mathbf{k}} \nonumber \\
&+ie_0\mathbf{E}(t)\cdot\sum_{\lambda,\mathbf{k}}\nabla_{\mathbf{k}} \lambda^{\dagger}_{\mathbf{k}}\lambda^{\mathstrut}_{\mathbf{k}} \nonumber \\
&-\sum_{\mathbf{k}}\left(\sum_{\mathbf{k}'} V_{\mathbf{k-k'}} p^*_{\mathbf{k'}} - \mathbf{d}_{\mathbf{k}}\cdot\mathbf{E}(t)\right)v^{\dagger}_{\mathbf{k}}c^{\mathstrut}_{\mathbf{k}} - \sum_{\mathbf{k}}\left(\sum_{\mathbf{k}'} V_{\mathbf{k-k'}} p_{\mathbf{k'}} - \mathbf{d}^*_{\mathbf{k}}\cdot\mathbf{E}(t)\right)c^{\dagger}_{\mathbf{k}}v^{\mathstrut}_{\mathbf{k}} \label{eq:HMF} \\
&=\sum_{\lambda,\mathbf{k}} \tilde{\varepsilon}_{\lambda,\mathbf{k}} \lambda^{\dagger}_{\mathbf{k}}\lambda^{\mathstrut}_{\mathbf{k}} - \sum_{\mathbf{k}}\Delta_{\mathbf{k}}\left( v^{\dagger}_{\mathbf{k}}c^{\mathstrut}_{\mathbf{k}} +  c^{\dagger}_{\mathbf{k}}v^{\mathstrut}_{\mathbf{k}}  \right) +ie_0\sum_{\lambda,\mathbf{k}}\mathbf{E}(t)\cdot(\nabla_{\mathbf{k}}\lambda^{\dagger}_{\mathbf{k}})\lambda^{\mathstrut}_{\mathbf{k}} + \hbar\sum_{\mathbf{k}}\Omega_{\mathbf{k}}(t)\left( v^{\dagger}_{\mathbf{k}}c^{\mathstrut}_{\mathbf{k}} + c^{\dagger}_{\mathbf{k}}v^{\mathstrut}_{\mathbf{k}}  \right) . \label{eq:Hamilton}
\end{align}
which reproduces the semiconductor Bloch equations in the Hartree-Fock limit, including effects as exciton formation or band gap renormalization. The single-particle energies $\tilde{\varepsilon}_{\lambda,\mathbf{k}}$ are renormalized by Hartree-Fock contributions of intra- and interlayer Coulomb interaction, which contain the carrier occupation $f_{\lambda,\mathbf{k}}=\langle \lambda^{\dagger}_{\mathbf{k}}\lambda^{\mathstrut}_{\mathbf{k}}\rangle$ in TMDC and molecule layer and $\Delta_{\mathbf{k}}=\sum_{\mathbf{k}'}V_{\mathbf{k-k'}}p_{\mathbf{k}'}$ accounts for the formation of bound excitons and corresponds to the built up of a macroscopic coherence in the EI state. The microscopic transitions are defined as $p_{\mathbf{k}}=\langle v^{\dagger}_{\mathbf{k}}c^{\mathstrut}_{\mathbf{k}}\rangle$ and $V_{\mathbf{k}}$ denotes the interlayer Coulomb potential \cite{Ovesen2019}. The last two terms include light-matter interaction consisting of intra- and interband transitions. The latter are determined by the Rabi frequency $\Omega_{\mathbf{k}}=\mathbf{d}_{\mathbf{k}}\cdot\mathbf{E}(t)/\hbar$ with electronic dipole moment $\mathbf{d}_{\mathbf{k}}$ and electric field $\mathbf{E}(t)$. In this manuscript, we use the band gap energy as parameter, tunable by a static external additional out-of-plane electric field via the Stark effect \cite{Lorchat2021}. This tuning enables different electronic phases of the heterostructure. Since the excitonic insulator phase results from a spontaneous formation of excitons, we first focus on the ground state calculation in absence of an external exciting optical field.

The Hamiltonian is then diagonalized with a Bogoliubov transformation. The new ground state is given by \cite{Jerome1967}
\begin{align}
|\tilde{\Psi}_{0}\rangle = \Pi_{\mathbf{k}}\left( u^*_{\mathbf{k}} -w^*_{\mathbf{k}}c^{\dagger}_{\mathbf{k}}v^{\mathstrut}_{\mathbf{k}} \right)|\Psi_0\rangle=\Pi_{\mathbf{k}} \alpha^{\dagger}_{\mathbf{k}}|0\rangle 
\end{align}
with $|\Psi_0\rangle=\Pi_{\mathbf{k}}v^{\dagger}_{\mathbf{k}}|0\rangle$ as the conventional semiconducting ground state constructed from the vacuum state $|0\rangle$. The coherence factors $|w_{\mathbf{k}}|^2$ and $|u_{\mathbf{k}}|^2$ describe the probabilities that the pair state is occupied or unoccupied, respectively. The diagonalization yields the coherence factors
\begin{align}
|u_{\mathbf{k}}|^2&=\frac{1}{2}\left(1+\frac{\Sigma_{\mathbf{k}}}{\sqrt{\Sigma_{\mathbf{k}}^2+\Delta_{\mathbf{k}}^2}}\right), \quad\text{and}\quad |w_{\mathbf{k}}|^2=\frac{1}{2}\left(1-\frac{\Sigma_{\mathbf{k}}}{\sqrt{\Sigma_{\mathbf{k}}^2+\Delta_{\mathbf{k}}^2}}\right) \;.
\end{align}
Finally, we define the Coulomb potential
\begin{align}
V_{\mathbf{q}}^{ll'}&=\frac{e_0^2}{2\epsilon_0A|\mathbf{q}|\epsilon^{ll'}_{\mathbf{q}}} ;\quad \epsilon_{\mathbf{q}}^{ll'}=\left\{\begin{array}{ll} \epsilon_{\mathbf{q}}, & l\neq l' \\
         \epsilon^i_{\mathbf{q}}, & l=l'\equiv i\end{array}\right. \label{eq:VCoul}
\end{align}
with $i=\{0,1\}$ for molecular and TMDC layer. The dielectric functions read
\begin{align}
\epsilon_{\mathbf{q}}=\kappa g^0_{|\mathbf{q}|}g^1_{|\mathbf{q}|}f_{|\mathbf{q}|} \quad \text{and}\quad  \epsilon^{i}_{\mathbf{q}}&=\frac{\kappa g^{1-i}_{|\mathbf{q}|}f_{|\mathbf{q}|}}{\cosh(\delta_{1-i}{|\mathbf{q}|}/2)h^i_{|\mathbf{q}|}} 
\end{align}
with the abbreviations
\begin{align}
f_{\mathbf{q}}&=1+\frac{1}{2}\left((\frac{\kappa_0}{\kappa}+\frac{\kappa}{\kappa_0})\tanh(\delta_0|\mathbf{q}|)+(\frac{\kappa_1}{\kappa}+\frac{\kappa}{\kappa_1})\tanh(\delta_1|\mathbf{q}|)+(\frac{\kappa_0}{\kappa_1}+\frac{\kappa_1}{\kappa_0})\tanh(d_0 |\mathbf{q}|)\tanh(\delta_1 |\mathbf{q}|)\right) \\
h_{\mathbf{q}}^i&=1+\frac{\kappa}{\kappa_i}\tanh(\delta_i|\mathbf{q}|)+\frac{\kappa}{\kappa_{1-i}}\tanh(\delta_{1-i}|\mathbf{q}|/2)+\frac{\kappa_i}{\kappa_{1-i}}\tanh(\delta_i|\mathbf{q}|)\tanh(\delta_{1-i}|\mathbf{q}|/2) \\
g_{\mathbf{q}}^i&=\frac{\cosh(\delta_i|\mathbf{q}|)}{\cosh(\delta_{1-i}|\mathbf{q}|/2)}\left(1+\frac{\kappa}{\kappa_i}\tanh(\delta_i|\mathbf{q}|/2)\right) .
\end{align}
The parameters are $\kappa_i=\sqrt{\epsilon_{\parallel}^i\epsilon_{\perp}^i}$ and $\kappa$ for the dielectric background, $\alpha_i=\sqrt{\epsilon^i_{\parallel}/\epsilon^i_{\perp}}$, $\delta_i=\alpha_id_i$ with the layer thickness $d_i$.

\section{Equations of motion}
The macroscopic polarization is defined as $\mathbf{P}(t)=-\delta H_{lm}/\delta\mathbf{E}(t)$ with the light-matter Hamiltonian $H_{lm}$. Already in excitonic basis, the macroscopic polarization reads
\begin{align}
\mathbf{P}(t)=-\sum_{\mu,\mathbf{k}}\left(f^{(0)}_{\mathbf{k}} \varphi_{\mu,\mathbf{k}}\mathbf{d}_{\mathbf{k}} + ie_0\varphi^*_{\mu,\mathbf{k}}\nabla_{\mathbf{k}}p^{(0)}_{\mathbf{k}}\right) P_{\mu} + ie_0\sum_{\lambda,\mathbf{k}}[\nabla_{\mathbf{k}}\lambda^{\dagger}_{\mathbf{k}}]\lambda^{\mathstrut}_{\mathbf{k}}
\end{align}
from the macroscopic polarization we can calculate the optical current
\begin{align}
\mathbf{j}(t)&=-\sum_{\mu,\mathbf{k}} \left( \mathbf{d}_{\mathbf{k}}\varphi_{\mu,\mathbf{k}} f^{(0)}_{\mathbf{k}} + ie_0\varphi_{\mu,\mathbf{k}}\nabla_{\mathbf{k}}p^{(0)}_{\mathbf{k}}\right)\frac{d}{dt}P_{\mu}(t)+e_0\sum_{\mathbf{k}}\mathbf{v}_{\mathbf{k}} F^{(1)}_{v,\mathbf{k}}(t) \label{eq:jt} \\
\mathbf{j}(\omega)&=-\sum_{\mu,\mathbf{k}} \left( \mathbf{d}_{\mathbf{k}}\varphi_{\mu,\mathbf{k}} f^{(0)}_{\mathbf{k}} + ie_0\varphi_{\mu,\mathbf{k}}\nabla_{\mathbf{k}}p^{(0)}_{\mathbf{k}}\right)i\omega P_{\mu}(\omega)+e_0\sum_{\mathbf{k}}\mathbf{v}_{\mathbf{k}} F^{(1)}_{v,\mathbf{k}}(\omega) \label{eq:jw}
\end{align}
where we introduced the particle velocity $\mathbf{v}_{\mathbf{k}}=\hbar\mathbf{k}/m$ and Fourier transformed to get from Eq. \eqref{eq:jt} to Eq. \eqref{eq:jw}. Since the molecule electrons are infinitely heavy they do not contribute to the current that only the valence band electrons $F_{v,\kp}^{(1)}$ are relevant. We derive the equations of motion for the optical excitations, which read
\begin{align}
i\hbar\frac{d}{dt}p_{\mathbf{k}} &= \left(2\Sigma_{\mathbf{k}} + ie_0\mathbf{E}(t)\cdot\nabla_{\mathbf{k}}\right) p_{\mathbf{k}} - \Delta_{\mathbf{k}}f_{\mathbf{k}} +\hbar\Omega_{\mathbf{k}}f_{\mathbf{k}} \\
i\hbar\frac{d}{dt}f_{\mathbf{k}} &= 2i\im\left(\Delta_{\mathbf{k}}p_{\mathbf{k}} \right) +e_0\mathbf{E}(t)\cdot\nabla_{\mathbf{k}}f_{\mathbf{k}}
\end{align}
with $2\Sigma_{\mathbf{k}}=\tilde{\varepsilon}_{c,\mathbf{k}}-\tilde{\varepsilon}_{v,\mathbf{k}}$ and inversion $f_{\mathbf{k}}=f_{v,\mathbf{k}}-f_{c,\mathbf{k}}$. Then, we expand the polarization and density into orders of the exciting electric field
\begin{align}
p_{\mathbf{k}}&=p^{(0)}_{\mathbf{k}}+p^{(1)}_{\mathbf{k}}+\mathcal{O}(2) \\
f_{\mathbf{k}}&=f^{(0)}_{\mathbf{k}}+f^{(1)}_{\mathbf{k}}+\mathcal{O}(2) .
\end{align}
Staying in the first order of the electric field the HIOS Bloch equation reads
\begin{align}
i\hbar\frac{d}{dt}p^{(1)}_{\mathbf{k}}&=2\Sigma^{(0)}_{\mathbf{k}}p^{(1)}_{\mathbf{k}} - \Delta^{(1)}_{\mathbf{k}}f^{(0)}_{\mathbf{k}} + 2\Sigma^{(1)}_{\mathbf{k}}p^{(0)}_{\mathbf{k}} - \Delta^{(0)}_{\mathbf{k}}f^{(1)}_{\mathbf{k}} +\hbar\Omega_{\mathbf{k}}f^{(0)}_{\mathbf{k}} + ie_0\mathbf{E}(t)\cdot\nabla_{\mathbf{k}}p^{(0)}_{\mathbf{k}} \label{eq:p} \\
i\hbar\frac{d}{dt}f^{(1)}_{\mathbf{k}}&=2i\Im\left(\Delta^{(0)}_{\mathbf{k}}p^{(1)}_{\mathbf{k}}+\Delta^{(1)}_{\mathbf{k}}p^{(0)}_{\mathbf{k}} \right)+ie_0\mathbf{E}(t)\cdot\nabla_{\mathbf{k}}f^{(0)}_{\mathbf{k}} .
\end{align}
Since $p^{(0)}_{\mathbf{k}}$ and $f^{(0)}_{\mathbf{k}}$ describe the ground state their dynamics vanish. The first two terms of Eq. \eqref{eq:p} correspond to the semiconductor Bloch equation. The second two terms are new and couple ground state distributions and excited quantities. The last two terms describe the optical excitation with interband source with included Pauli blocking \cite{Selig2019} and intraband source, respectively. The coupling induced by the third and fourth term  can be resolved via the transformation \cite{Stroucken2015}
\begin{align}
P^{(1)}_{\mathbf{k}} &= \frac{E_{\mathbf{k}}+\Sigma^{(0)}_{\mathbf{k}}}{E_{\mathbf{k}}} p^{(1)}_{\mathbf{k}} - \frac{E_{\mathbf{k}}-\Sigma^{(0)}_{\mathbf{k}}}{E_{\mathbf{k}}} p^{*(1)}_{\mathbf{k}} - \frac{\Delta^{(0)}_{\mathbf{k}}}{E_{\mathbf{k}}} f^{(1)}_{\mathbf{k}} \\
F^{(1)}_{\mathbf{k}} &= \frac{\Delta^{(0)}_{\mathbf{k}}}{E_{\mathbf{k}}} \left(p^{(1)}_{\mathbf{k}}+p^{*(1)}_{\mathbf{k}} \right) + \frac{\Sigma^{(0)}_{\mathbf{k}}}{E_{\mathbf{k}}} f^{(1)}_{\mathbf{k}} .
\end{align}
We obtain
\begin{align}
i\hbar\frac{d}{dt}P^{(1)}_{\mathbf{k}} &= 2E_{\mathbf{k}}P^{(1)}_{\mathbf{k}} - \sum_{\mathbf{k'}}V_{\mathbf{k-k'}}P^{(1)}_{\mathbf{k}'} +\hbar\Omega_{\mathbf{k}}f^{(0)}_{\mathbf{k}} + ie_0\mathbf{E}(t) \cdot\nabla_{\mathbf{k}} p^{(0)}_{\mathbf{k}} \\
i\hbar\frac{d}{dt}F^{(1)}_{\mathbf{k}} &= ie_0\mathbf{E}(t)\cdot\nabla_{\mathbf{k}}f^{(0)}_{\mathbf{k}}
\end{align}
We can identify the Bogoliubov-Wannier equation
\begin{align}
2E_{\mathbf{k}}\varphi_{\mu,\mathbf{k}}-\sum_{\mathbf{k'}}V_{\mathbf{k-k'}}\varphi_{\mu,\mathbf{k}'} = E_{\mu}\varphi_{\mu,\mathbf{k}}
\end{align}
which yields the excitonic equations
\begin{align}
i\hbar\frac{d}{dt}P^{(1)}_{\mu}&=E_{\mu}P^{(1)}_{\mu} + \sum_{\mathbf{k}} f^{(0)}_{\mathbf{k}}\varphi^*_{\mu,\mathbf{k}}\hbar\Omega_{\mathbf{k}} + ie_0\mathbf{E}(t)\cdot\sum_{\mathbf{k}} \varphi^*_{\mu,\mathbf{k}}\nabla_{\mathbf{k}}p^{(0)}_{\mathbf{k}} \label{eq:P} \\
i\hbar\frac{d}{dt}F^{(1)}_{\mathbf{k}} &= ie_0\mathbf{E}(t)\cdot\nabla_{\mathbf{k}}f^{(0)}_{\mathbf{k}} \label{eq:F}
\end{align}
Equation \eqref{eq:P} can be Fourier transformed an inserted into Eq. \eqref{eq:jw}. For the occupations we Fourier transform Eq. \eqref{eq:F} with phenomenological dephasing and insert into Eq. \eqref{eq:jw}, which yields for the last term
\begin{align}
\mathbf{j}(\omega)=-e_0\sum_{\mathbf{k}}\mathbf{v}_{\mathbf{k}}\frac{ie_0\mathbf{E}(t)\nabla_{\mathbf{k}}f^{(0)}_{v,\mathbf{k}}}{\hbar\omega+i\gamma}=\left(-\frac{ie_0^2}{\hbar}\sum_{\mathbf{k}}\frac{\mathbf{v}_{\mathbf{k}}\otimes\nabla_{\mathbf{k}}f^{(0)}_{v,\mathbf{k}}}{\omega+i\gamma/\hbar}\right)\cdot\mathbf{E}(t) \label{eq:DrudeTensor}
\end{align}
Comparing with the definition of the current $\mathbf{j}(\omega)=-i\omega\epsilon_0\chi(\omega)\mathbf{E}(\omega)$ we can identify the scalar susceptibility
\begin{align}
\chi(\omega)=-\frac{1}{\epsilon_0}\sum_{\mu}\frac{\mathbf{d}_{\mu}\otimes\mathbf{d}_{\mu} + \mathbf{j}_{\mu}\otimes\mathbf{j}_{\mu}}{\hbar\omega-E_{\mu}+i\gamma} +\frac{e_0^2}{\epsilon_0\hbar}\sum_{\mathbf{k}}\frac{\mathbf{v}_{\mathbf{k}}\otimes\nabla_{\mathbf{k}}f^{(0)}_{v,\mathbf{k}}}{\omega^2+i\omega\gamma/\hbar} .
\end{align}
In case that the electronic phase has a valence occupation $f^{(0)}_{v,\mathbf{k}}$ with Fermi edge, we can partially integrate the last term in Eq. \eqref{eq:DrudeTensor} and find
\begin{align}
\left(-\frac{ie_0^2}{\hbar}\sum_{\mathbf{k}}\frac{\mathbf{v}_{\mathbf{k}}\otimes\nabla_{\mathbf{k}}f^{(0)}_{v,\mathbf{k}}}{\omega+i\gamma/\hbar}\right)\cdot\mathbf{E}(t)=\frac{ie_0^2}{m}n_{el}\frac{1}{\omega+i\gamma/\hbar}\mathbf{E}(\omega)
\end{align}
where defined the carrier number $n_{el}=\sum_{\mathbf{k}}f^{(0)}_{v,\mathbf{k}}$. Again comparing with the definition of the optic current and assuming a perpendicular excitation we find the scalar susceptibility
\begin{align}
\chi(\omega)=-\frac{1}{\epsilon_0}\sum_{\mu}\frac{|d_{\mu}|^2 + |j_{\mu}|^2}{\hbar\omega-E_{\mu}+i\gamma} + \frac{\omega^2_{pl}}{\omega^2+i\omega\gamma/\hbar}
\end{align}
with the plasma frequency $\omega^2_{pl}=e_0^2n_{el}/\epsilon_0m$.

\section{Optical selection rules of excitonic insulator}
Investigating the gap equation Eq. (1) of the main text we see that it corresponds to the Bogoliubov-Wannier equation with vanishing exciton binding energy. This suggests that also the ground state polarization could be projected onto the wave function acting as solution for $E_{\mu}=0$: $p^{(0)}_{\mathbf{k}}=\sum_{\nu,E_{\nu}=0}\varphi_{\nu,\mathbf{k}}p^{(0)}_{\nu}$. Then the momentum-gradient in the intraband matrix element acts onto the wave function. For a $s$-type ground state the angular derivative vanishes. Together with an analytical treatment of the angle-sum we obtain for the intraband source
\begin{align}
j_{\mu}=e_0\pi\mathbf{e}\cdot\sum_{\nu,k}\varphi_{\mu,k}^*\partial_k\varphi_{\nu,k}
\left(
\begin{array}{c}
1  \\
\pm i 
\end{array}
\right)
\end{align}
where the sign stands for $\mu=p_+$ and $\mu=p_-$ final states. These two final states exhibits a circular dichroismic selection rule, comparable to KK- and K$^-$K$^{-}$-excitons in monolayer TMDCs.

\bibliographystyle{apsrev4-1}
%

\end{document}